# Imaging defects in two-dimensional crystals by convergent beam electron diffraction

*Tatiana Latychevskaia[1], Pengru Huang[2], Kostya S. Novoselov[2,3,4,5]*

[1]*Paul Scherrer Institute, 5232 Villigen, Switzerland*

[2]*Institute for Functional Intelligent Materials, National University of Singapore, Singapore, 117544, Singapore*

[3]*Centre for Advanced 2D Materials, National University of Singapore, 117546 Singapore*

[4]*Chongqing 2D Materials Institute, Liangjiang New Area, Chongqing, 400714, China*

[5]*National Graphene Institute, University of Manchester, Oxford Road, Manchester, M13 9PL, UK*

## Abstract

Convergent beam electron diffraction (CBED), recently demonstrated on two-dimensional (2D) materials, offers a number of interesting applications such as imaging atomic in- and out-of plane shifts, interlayer distances, and individual adsorbates. In this study, we show how CBED allows for atomic-precision imaging of individual defects in 2D materials using one single-shot intensity measurement. In combination with structural calculations using density functional theory, we present simulated CBED patterns for various defects in graphene, each of which exhibits a unique fingerprint distribution. We also show how atomic positions, including the individual atomic defects in graphene, can be reconstructed by iterative phase retrieval from a single CBED pattern.

## Main

An understanding of the internal structure of defects lies at the core of many technologies, from single-atom catalysis[1-3] to the formation of coloured centres in crystals[4]. The local atomic-sized defects in graphene define the transport properties of this material[5]. At the same time, transmission electron microscopy (TEM) remains the only technique which can reliably deliver information about the imperfections in both metallic and insulating solids at the atomic scale. The quality and resolution of the images of defects obtained using conventional TEM relies strongly on the performance of the lenses[6], which are effectively used as a Fourier transform (FT) apparatus to convert an electron beam diffracted on a crystal into a real-space image. Despite the availability of electron microscopes since the 1930s, it is only since the 2000s that progress in aberration-corrected electron optics[7-9] has allowed for atomic-resolution images of crystalline samples to be obtained[6, 10]. Scanning transmission electron microscopy (STEM) is another alternative that allows for atomic-resolution images of a sample[11]. Recent progress in the imaging of 2D materials has been made by applying scanning techniques such as 4D-STEM[12, 13], in which a 2D diffraction pattern is acquired at each beam position while scanning the sample, and the obtained 4D dataset is analysed by applying an algorithm developed for ptychography. Through the use of this electron ptychography technique, images of bilayer $MoS_2$ have been obtained with a resolution of 0.4 Å[14].

There is, however, an alternative methodology for recovering real-space information about an object from a *single-shot* intensity measurement of the diffracted electron beams, which does not require high-quality lenses or scanning: we can utilise reconstruction algorithms that can recover an image of the sample directly from the diffracted waves. Since the phase information in the diffraction pattern is lost during projection on the screen, iterative phase retrieval algorithms can be used, and particular constraints must be introduced to reconstruct the sample distribution[15].

In this paper, we demonstrate that the use of a conventional electron microscope in convergent beam electron diffraction (CBED) mode allows us to obtain "fingerprint" CBED patterns of the individual defects in graphene. CBED has previously been applied to study two-dimensional (2D) materials[16-19], resulting in a number of interesting applications such as the imaging of atomic in- and out-of plane shifts[16], interlayer distances[16, 18], and individual adsorbates[19]. In the present study, we demonstrate that CBED patterns and particular constraints allow for the extraction of atomic information from CBED images, and can be applied to image individual defects in 2D materials. This high-quality recovery of real-space information about the object eliminates the need to use expensive aberration-corrected optics or scanning procedures. The use of a single-shot intensity acquisition regime offers the possibility of studying fast dynamic processes such as adsorbate dynamics[20-22].

# Results

## CBED patterns of graphene with individual defects

A schematic diagram of a CBED experiment is shown in Fig. 1(a), and a single defect in graphene is illustrated in Fig. 1(b). The semi-convergence angle $\alpha$ is selected in such a way that the individual CBED discs do not overlap, and the interference patterns in the individual CBED discs can be analysed separately. All of the CBED patterns presented in this study were simulated for electrons of 80 keV and $\alpha$ = 7 mrad. By changing the defocus value $\Delta f$, the diameter of the probed region can be varied as $D = 2\,\Delta f \tan\alpha$. The distributions of the incident wavefront at $\Delta f = -0.1$ μm and $\Delta f = -1$ μm are shown in Figs. 1(c) and (d), respectively, and the corresponding CBED patterns are shown in Figs. 1(e) and (f). The diameter of the probed area is about 0.7 nm (counted as the distance between the first minima) and 14 nm for defocus values of $\Delta f = -0.1$ μm and $\Delta f = -1$ μm, respectively. A defocus value of $\Delta f$ = −1 μm is optimal for the identification of interference patterns due to an atomic defect appearing in individual CBED discs, as shown by the example of a single defect visualised in a CBED pattern in Figs. 1(f) and (g). At this defocus value, the extent of the interference pattern due to an atomic defect is completely within the CBED disc, as shown in Fig. 1(g). The atomic lattice defects cause a localised interference pattern that can be described as concentric rings, unlike other lattice defects such as stacking faults or ripples, which create extended interference patterns in CBED discs, as previously illustrated in ref [16].

Various atomic defects were considered in our simulations, and the resulting CBED patterns are shown in Fig. 2. It is widely accepted that DFT calculations allow for very accurate geometry predictions[23]. In particular, they can closely reproduce the experimentally determined defect structures in 2D materials[24]. In this study, the equilibrium atomic geometries of the defects, used as inputs for the CBED simulations, were obtained from DFT calculations, as explained in the section entitled "Methods: DFT atomic coordinates". The corresponding CBED patterns simulated at a defocus value of $\Delta f$ = −1 μm are shown in Fig. 2, and the simulation procedure is described in the section entitled "Methods: Simulation of CBED patterns". The interference patterns in the individual CBED discs exhibit concentric fringes. When the defect is positioned off-centre in the probed region, the interference pattern in each CBED disc appears at the same offset from the centre, as illustrated in Figs. 2(h) and (p). The contrast of the interference patterns in the individual CBED spots, calculated as $C = \left(I_{\max} - I_{\min}\right) / \left(I_{\max} + I_{\min}\right)$, is almost zero for the zero-order CBED spot, and increases linearly with the order for the higher-order CBED spots. For example, for a double defect (Fig. 2(a)), the contrast is 0.03, 0.24±0.09, and 0.42±0.10, for the zero-, first-, and second-order CBED spots, respectively. The contrast is sufficient for recognition of a characteristic interference pattern due to a defect already in a first-order CBED disc. From Fig. 2, it is evident that each defect exhibits a distinct "fingerprint" CBED pattern, which in principle can be used to directly assign the defect. Examples of experimental CBED patterns of monolayer hexagonal boron nitride with adsorbates were previously reported ref.[19]. It should be noted that a direct manifestation of an atomic defect through a "fingerprint" interference pattern cannot be obtained using a conventional diffraction pattern, in which sharp diffraction peaks are formed by the superposition of all of the diffracted waves from the entire probed area.

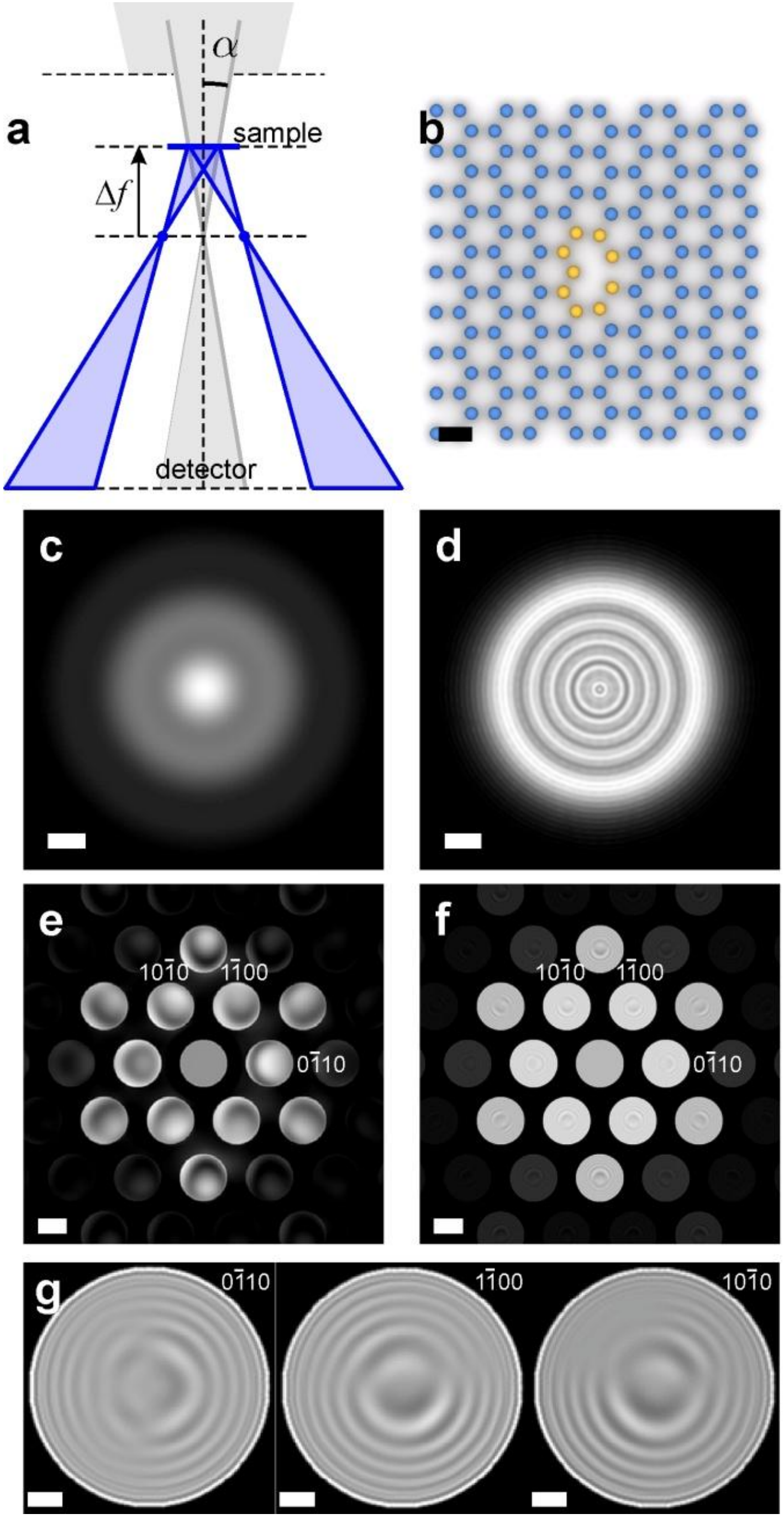

Fig. 1. Convergent beam electron diffraction (CBED) imaging of atomic defects in graphene. (a) Scheme used for CBED imaging in a transmission electron microscope. (b) Sketch showing a single defect in graphene. (c) and (d) Amplitude distributions of the probing wavefront when imaging for defocus values of $\Delta f$ = −0.1 and −1 μm, respectively. (e) and (d) CBED patterns of single defect in graphene simulated at $\Delta f$ = −0.1 and −1 μm, respectively, where the intensity of the zero-order CBED disc is reduced by a factor of 10E+4. (e) Magnified individual CBED discs in the CBED pattern at $\Delta f$ = −1 μm. The scale bars in (b) and (c) represent 2 Å, (d) and (g) 2 nm, (e) and (f) 2 1/nm.

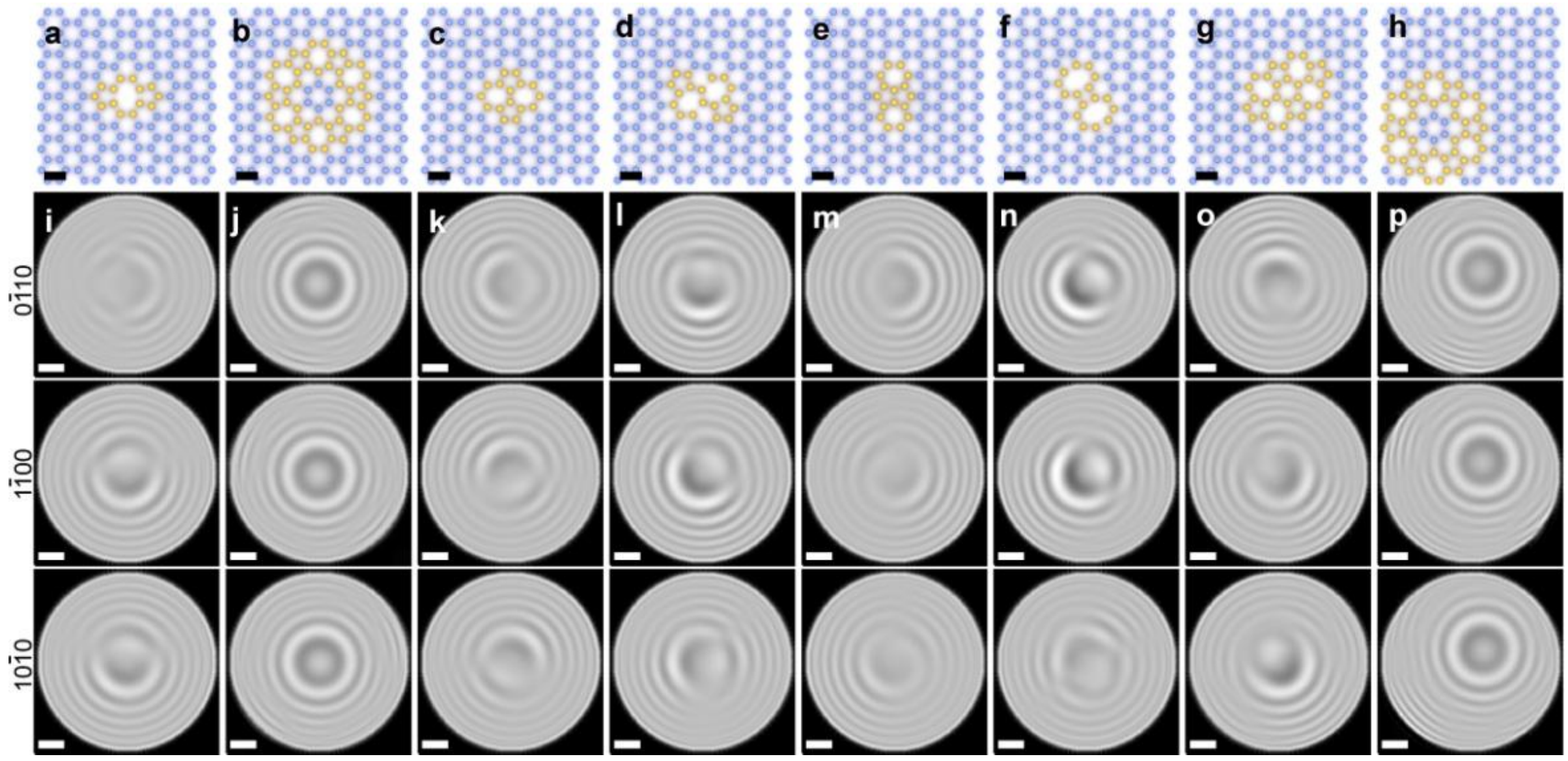

Fig. 2. Various atomic defects in graphene: (a) double vacancy (5-8-5), (b) miniature grain boundary formed by the rotation of a core of seven hexagons, (c) Stone-Wales defect, (d) 5-7-7-5 defect reconstructed from the double vacancy, (e) inverse Stone-Wales defect formed by the incorporation of two carbon adatoms into the lattice, (f) tetravacancy (5-8-4-8-5) and (g) 555-6-777 defect reconstructed from the double vacancy, (h) the same defect as in (b) but shifted downwards towards the left, and (i)-(p) their corresponding CBED patterns simulated at $\Delta f = -1$ μm, shown as individual CBED spots $\left(0\bar{1}10\right)$ (top row), $\left(1\bar{1}00\right)$ (middle row) and $\left(10\bar{1}0\right)$ (bottom row). The scalebars in (a)-(h) represent 2 Å, (i)-(p) 2 nm. The range of intensity values (i.e. the minimal and maximal intensity values) are the same for all images in (i)-(p).

## Reconstruction of individual defects

The atomic structure of a sample can be reconstructed from its CBED pattern by applying iterative phase retrieval methods[25-29], provided that CBED patterns are acquired under geometric conditions that allow for correct sampling. The digitally sampled distributions in the detector and sample planes are related via an FT, and hence the pixel sizes in the sample plane ($\Delta_0$) and in the detector (CBED pattern) plane ($\Delta_k$) are related as $\Delta_0\Delta_k = 2\pi/N$, where $N \times N$ is the number of pixels. The conditions that allow for reconstruction are as follows. (i) The number of pixels $N$ and the pixel size in the sample plane $\Delta_0$ should be selected such that there are several pixels between the closest positioned atoms, to allow the atoms to be resolved from one another. (ii) The defocus distance $\Delta f$ must be selected such that the probed region is less than or equal to twice the reconstructed area (entire field of view) in order to fulfil the oversampling condition. This will ensure that the non-probed region will contain pixels with zero values in the sample plane. The number of these pixels should be larger than the number of pixels with unknown values, so that the system of equations has a unique solution. The diameter of the probed region is approximately given by the diameter of the probing beam, which linearly depends on the defocus value as $D = 2\,\Delta f \tan\alpha$. (iii) After the number of pixels $N$ and the pixel size in the sample plane $\Delta_0$ have been selected under condition (i), the $k$-range in the detector plane and the pixel size in the detector plane should be selected such that the condition $\Delta_0\Delta_k = 2\pi/N$ is fulfilled.

A CBED pattern of graphene layer with a single defect (Fig. 1(b)) was simulated under the following parameters. The entire field of view was 20 nm × 20 nm, and was sampled with 2048 × 2048 pixels; hence, the pixel size in the sample plane was $\Delta_0$ = 9.8 pm, with 14 pixels between the closest carbon atoms. From the condition $\Delta_0\Delta_k = 2\pi/N$, we can obtain the pixel size in the CBED pattern plane as $\Delta_k$ = 5E+7 1/m. The defocus value was set to $\Delta f$ = −0.1 μm, giving an area of approximately 1.4 nm in diameter at the semi-convergence angle $\alpha$ = 7 mrad. However, due to diffraction effects, the probing beam is not a sharp-edged disc with diameter of 1.4 nm, but is a distribution arising from diffraction on an aperture, as shown in Fig. 1(c). The extent of the probing beam is much smaller than the field of view of 20 × 20 nm. Thus, the oversampling condition is fulfilled. The simulated CBED pattern is shown in Fig. 1(e).

A CBED pattern is an intensity distribution, meaning that the phase of the wavefront in the detector plane is lost. In order to recover the atomic positions, the phase information needs to be restored, which can be done by applying iterative phase retrieval algorithms[29]. The CBED pattern shown in Fig. 1(e) was reconstructed by applying an iterative phase retrieval procedure, as explained in the section entitled “Methods: Iterative phase retrieval”. The atomic positions reconstructed after 100 iterations perfectly match the atomic distribution in the sample (Fig. 3). The intensity of the central spot was assumed to be blocked by a beam stop and unavailable, which is typical in an experiment. It should be noted that blocking of the central disc in a CBED pattern is not only a typical experimental arrangement but was also found to be necessary to achieve artefact-free sample reconstruction. When the central disc in the CBED pattern is available, its inverse FT gives intense concentric rings in the first iteration, which are superimposed onto the reconstructed sample distribution. These rings do not exactly match the distribution of the incident wavefront, and hence a division by the incident wavefront does not remove them. The obtained transmission function of the sample therefore still contains these artefact rings, and the amplitude and phase distributions cannot be extracted accurately. As the iterations progress, the concentric ring artefacts in the sample plane do not fade away; even after hundreds of iterations, these artefact rings are still present in the sample plane and the phase distribution of the sample is not correctly retrieved. In contrast, when the central disc in the CBED pattern is set to zero in the first iteration, the reconstructed sample amplitude and phase distributions are not contaminated by the artefact rings and the constraint can be applied to the phase distribution, resulting in improvements in the phase distribution with each iteration. As the iterations progress, the signal in the central CBED disc is gradually restored to its original values. Since the iterative procedure implies the use of an FT, and the distribution in the sample plane is described by a transmission function exp(*i**phase) (where the phase is unknown), the constant component of the transmission function guarantees that the zero-order spot will eventually be restored.

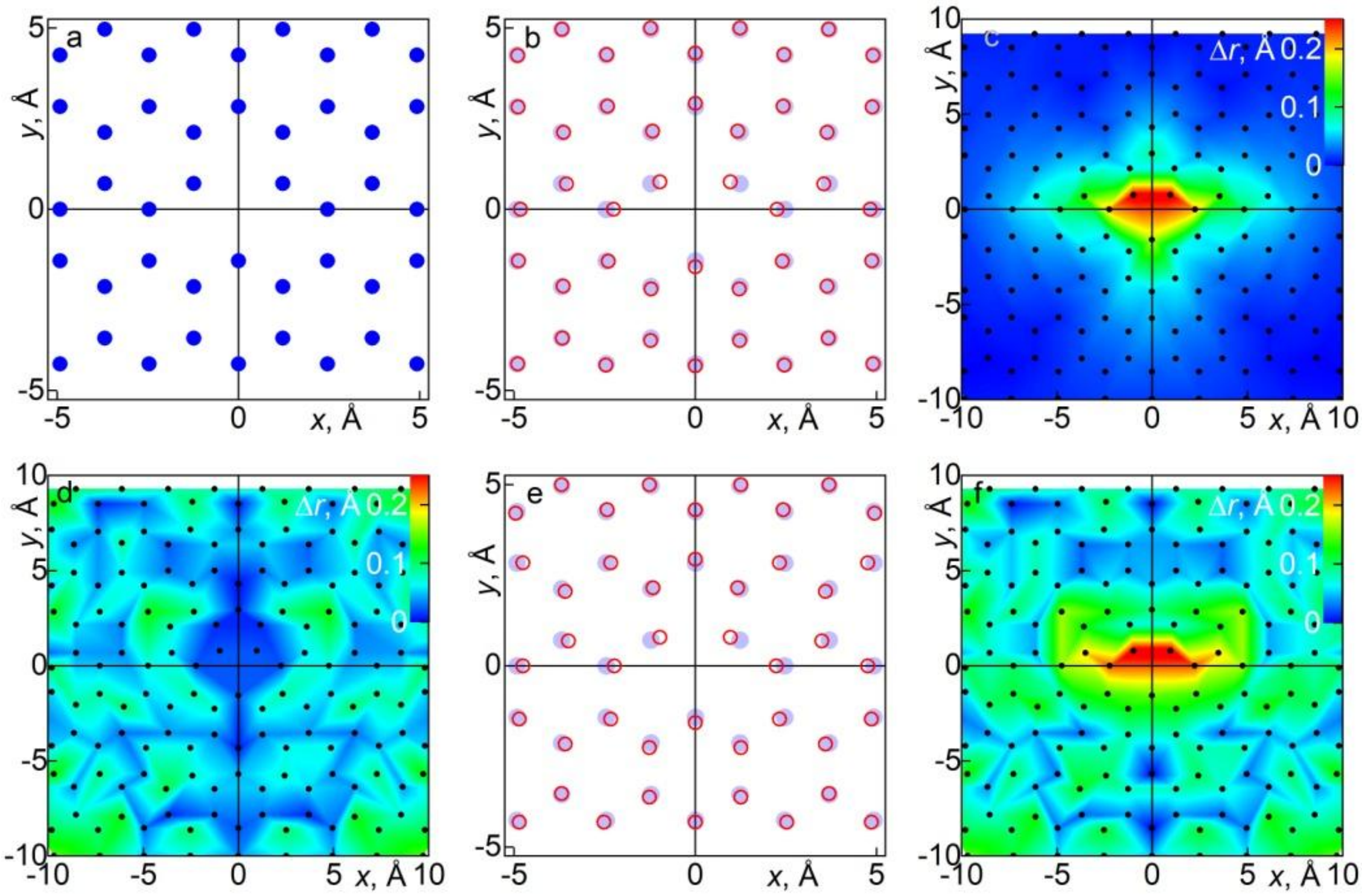

Fig. 3. Atomic displacements in graphene with a single defect reconstructed by iterative phase retrieval. (a) 2D distribution of atoms in a lattice without relaxation. (b) Relaxed positions of atoms around a defect (red circles; blue circles represent the same positions as shown in (a)). (c) Difference between the atomic positions in the relaxed lattice (red circles in (b)) and lattice without relaxation (blue circles in (b)). (d) Difference between the atomic positions in the relaxed lattice (red circles in (b)) and the reconstructed lattice (red circles in (e)). (e) Reconstructed positions of the atoms around a defect (red circles; blue circles represent the same positions as shown in (a)). (f) Difference between the atomic positions in the reconstructed lattice and the lattice without relaxation (difference between the atomic positions in (e)).

The precision of the reconstructed atomic positions is visualised in Fig. 3. The atomic shifts due to lattice relaxation reach 0.25 Å (Figs. 3(a) and (b)). It is evident that the atomic distribution without a single atom is correctly retrieved (Figs. 3(c)-(e)). The reconstructed atomic coordinates correctly reproduce the positions of the atoms in the relaxed lattice, with shifts due to relaxation of up to 0.25 Å (Fig. 3(f)). The maximal deviation of the reconstructed coordinates from the correct coordinates is 0.1 Å, and this originates not from the iterative phase retrieval but from the finite sampling of the transmission function of the atomic distribution and the finite pixel size of 9.8 pm. Finer sampling could reduce the error, although at the cost of increased computational time.

## Discussion

In the above, we demonstrated how CBED simulations, in combination with DFT structural predictions, enable the generation of unique fingerprints of most common defects in graphene. Our DFT simulations did not show out-of-plane distortions for the investigated defects, but we can assume that these can occur in general. The sensitivity of the contrast in CBED interference pattern is much lower for an out-of-plane atomic shift than for an in-plane shift. The phase shifts for out-of plane shift $\Delta z$ and in-plane shift $\Delta x$ are given by $\Delta\varphi_z \approx \Delta z \frac{\pi}{\lambda}\sin^2\vartheta$ and $\Delta\varphi_x = \Delta x \frac{2\pi}{\lambda}\sin\vartheta$, respectively[16]. A phase shift of 0.1 rad can be easily picked up by the interference pattern in a CBED disc, and can be created by $\Delta z$ = 0.35 nm or $\Delta x$ = 3.39 pm. Multiple defects can occur at a short distance from one another, and in this case, the resulting interference patterns in the CBED discs will appear as an in-line hologram of the defects, meaning that the complex-valued wavefronts rather than the intensities will be added coherently. The extent of this "overlap" between the complex-valued waves can be selected by choosing the size of the imaged area, that is, by regulating the defocus distance. Our DFT simulations suggest that the strain decays at the distance of few inter-atomic spacings.

We have also provided an outline of an iterative phase retrieval algorithm that allows for the recovery of the exact atomic positions in the monolayer graphene sample. Our iterative phase retrieval algorithm was modified from conventional phase retrieval algorithms in order to optimize the convergence by two distinct features: by removing the central part of the CBED pattern, and by using a perfect lattice as an initial guess. The modified algorithm has high stability and quick convergence even for noisy (SNR = 10) data, which reflects the most typical experimental conditions. For very low signal-to-noise ratio (SNR) imaging techniques, such as time-resolved methods, the algorithms described here may require additional modifications to tolerate low SNR. The iterative phase retrieval reconstruction approach does not require any *a priori* information about the amounts or shapes of individual defects, and although it is demonstrated here for the reconstruction of individual defects, the algorithm should work equally well for multiple defects in the field of view. These techniques may pave the way for the identification of atomic defects in 2D materials in a machinery manner. CBED provides a spatial precision of single atoms on the range of a few to a one hundreds of nanometres.

# Methods

## DFT atomic coordinates

The equilibrium geometries of our defect structures were obtained via DFT calculations using the PBE functional, as implemented in the Vienna Ab Initio Simulation Package (VASP 5.4)[30-32]. The valence electron configuration for the C elements is ($2s^2 2p^2$). The interaction between the valence electrons and ionic cores is described within the projector augmented (PAW) approach with a plane-wave energy cutoff of 500 eV[33]. Defect calculations were performed using an ~448-atom rectangular supercell of size (2.95 × 2.98) nm, enabling the position relaxation of atoms far from the defect centre to be less than a few pm. The DFT relaxed defect structures, as the central region, were embedded into a pristine graphene framework of size ~(60 × 60) nm and used as sample coordinates for CBED simulations. The Brillouin zone was sampled using a (1 × 1 × 1) Monkhorst-Pack grid. A 15 Å vacuum space was used to avoid interactions between neighbouring layers. In the structural energy minimization process, the atomic coordinates were allowed to relax until the forces on all the atoms were less than 0.01 eV/Å. All of the computational work for this article was performed using the resources of the National Supercomputing Centre, Singapore (https://www.nscc.sg).

## Simulation of CBED patterns

**Probing wave distribution.** The probing wave distribution $\psi_0(\vec{r})$ was modelled by simulating the diffraction of a spherical wavefront on a limiting virtual aperture (second condenser aperture) positioned at a plane $\vec{r}_0$:

$$\psi_0(\vec{r}) = \iint a(\vec{r}_0) \frac{\exp(-ikr_0)}{r_0} \frac{\exp(ik|\vec{r}_0 - \vec{r}|)}{|\vec{r}_0 - \vec{r}|} \mathrm{d}\vec{r}_0, \tag{1}$$

where $a(\vec{r}_0)$ is the aperture binary function. In the simulations, the integral transform in Eq. (1) was calculated directly, without applying a fast Fourier transform (FFT), as

$$\psi_0(p,q) = \sum_{m,n=-N/2}^{N/2-1} a(m,n) \frac{\exp\left(-ik\sqrt{(m\Delta_a)^2 + (n\Delta_a)^2}\right)}{\sqrt{(m\Delta_a)^2 + (n\Delta_a)^2}} \frac{\exp\left(ik\sqrt{(m\Delta_a - p\Delta)^2 + (n\Delta_a - q\Delta)^2}\right)}{\sqrt{(m\Delta_a - p\Delta)^2 + (n\Delta_a - q\Delta)^2}},$$

where *m*, *n*, *p*, *q* are the pixel indices in the aperture (*m*, *n*) and the sample (*p*, *q*) planes, and $\Delta_a$ and $\Delta$ are the pixel sizes in the aperture and the sample planes, respectively.

The semi-convergence angle of $\alpha$ = 7 mrad was achieved by selecting a virtual exit aperture of diameter 640 nm at a distance of 45 μm from the virtual source plane.

**CBED patterns of monolayer samples.** The input data consisted of an array of the coordinates of all the atoms $(x_n, y_n)$. The transmission function of a monolayer was calculated as:

$$t(x,y) = \exp\left[i\sigma v_z(x,y) \otimes l(x,y)\right], \tag{2}$$

where $v_z(x,y)$ is the projected potential of an individual atom, $l(x,y)$ is the function describing positions of the atoms in the lattice, and $\otimes$ denotes convolution. The projected potential of a single carbon atom was simulated in the form:

$$v_z(r) = 4\pi^2 a_0 e \sum_{i=1}^{3} a_i K_0\left(2\pi r\sqrt{b_i}\right) + 2\pi a_0 e \sum_{n=1}^{3} \frac{c_i}{d_i} \exp\left(-\pi^2 r^2 / d_i\right),$$

where $r = \sqrt{x^2 + y^2}$, $a_0$ is the Bohr radius, $e$ is the elementary charge, $K_0(\ldots)$ is the modified Bessel function, and $a_i, b_i, c_i, d_i$ are parameters that depend on the chemical origin of the atoms and are tabulated elsewhere[34]. In $v_z(r)$, the singularity at $r = 0$ is replaced by the value of $v_z(r)$ at *r* = 0.1 Å. The convolution $v_z(x,y) \otimes l(x,y)$ in Eq. (2) is calculated as $\mathrm{FT}^{-1}\left\{\mathrm{FT}\left[v_z(x,y)\right]\mathrm{FT}\left[l(x,y)\right]\right\}$. $\mathrm{FT}\left[l(x,y)\right]$ is simulated without applying an FFT in order to avoid artefacts associated with FFT, and is calculated as $\mathrm{FT}\left[l(x,y)\right] = \sum_n \exp\left[-i\left(k_x x_n + k_y y_n\right)\right]$, where $(x_n, y_n)$ are the exact atomic coordinates, not pixels. $\mathrm{FT}^{-1}\left\{\mathrm{FT}\left[v_z(x,y)\right]\mathrm{FT}\left[l(x,y)\right]\right\}$ is calculated by applying an inverse FFT to the product of $\mathrm{FT}\left[v_z(x,y)\right]$ and $\mathrm{FT}\left[l(x,y)\right]$.

The exit wave after passing through the sample is given by the product of the incident wave $\psi_0(x,y)$ and the transmission function of the sample $u(x,y) = \psi_0(x,y)t(x,y)$. The CBED pattern is then simulated as the square of the amplitude of the FT of the exit wave.

**Iterative phase retrieval.** The iterative phase retrieval was done by performing wavefront propagation back and forth between the sample and detector planes, using an FT and inverse FT, respectively.

(i) In the first iteration, the initial sample distribution was modelled as an incident wavefront $\psi_0(x,y)$ multiplied by the transmission function of a defect-free graphene lattice $t_0(x,y)$: $u(x,y)=\psi_0(x,y)t_0(x,y)$.

(ii) The FT of $u(x,y)$ was calculated to give the distribution of the wavefront in the far field: $\mathrm{FT}\left[u(x,y)\right]=U(X,Y)$. The amplitude of $U(X,Y)$ was replaced with the measured amplitude, giving $U'(X,Y)$.

(iii) The inverse FT of $U'(X,Y)$ was calculated to give the distribution of the wavefront in the sample plane $u(x,y)$. Constraints were applied to $u(x,y)$, giving $u'(x,y)$.

(iv) The iterative loop was restarted at (i).

During the iterative phase retrieval, the following constraints were applied. In the sample plane, the extracted phase distribution was set to be positive by setting the negative values to zero. In addition, a support constraint was applied in form of a mask 10 nm in diameter, where the values of the transmission function outside the mask were set to zero. In the detector plane in the central region, where no measured amplitudes were available, the amplitudes were updated in each iteration.

The progress of the reconstruction process can be monitored by the evaluating the error at the *k*-th iteration as[27] $E_k=\left\{\frac{\sum_{X,Y}\left[\left|U_k(X,Y)\right|-\left|U'(X,Y)\right|\right]^2}{\sum_{X,Y}\left|U'(X,Y)\right|^2}\right\}^{1/2}$. The iterative reconstruction can be stopped when the error stagnates, typically after tens of iterations.

## Author contributions

T. L. and K. N. initiated the research and analysed the data, P. H. designed the models and performed the DFT simulations, T. L. performed the diffraction simulations, all authors discussed the results and contributed to the final manuscript.

## Acknowledgements

The authors acknowledge support by the Ministry of Education (Singapore) through the Research Centre of Excellence program (Award EDUN C-33-18-279-V12, Institute for Functional Intelligent Materials), EU Graphene Flagship Program (contract CNECTICT-604391), European Research Council Synergy Grant Hetero2D (contract 319277), European Research Council Starting Grant EvoluTEM (contract 715502), the Royal Society (UK, grant number RSRP\R\190000), EPSRC grants EP/S019367/1, EP/P026850/1 and EP/N010345/1, EP/P009050/1, EP/S021531/1, FLAG-ERA project TRANS2DTMD, Swiss National Foundation Research Grant 200021_197107.